\documentstyle[aps,epsf]{revtex}

\begin{document}
\draft
\preprint{HEP/123-qed}


\newlength{\defitemindent} \setlength{\defitemindent}{.25in}
\newcommand{\deflabel}[1]{\hspace{\defitemindent}\bf #1\hfill}
\newenvironment{deflist}[1]%
	{\begin{list}{}
		{\itemsep=10pt \parsep=5pt \topsep=0pt \parskip=10pt
		\settowidth{\labelwidth}{\hspace{\defitemindent}\bf #1}%
		\setlength{\leftmargin}{\labelwidth}%
		\addtolength{\leftmargin}{\labelsep}%
		\renewcommand{\makelabel}{\deflabel}}}%
	{\end{list}}%


\newif\ifraisedcite
	\def\raisedcitations{\raisedcitetrue}
	\def\noraisedcitations{\raisedcitefalse}
	\noraisedcitations
\makeatletter
	\def\@cite#1#2{%
		\ifraisedcite\raisebox{.8ex}%
			{\scriptsize [{#1\if@tempswa,#2\fi}]}%
		\else [{#1\if@tempswa,#2\fi}]%
		\fi}
\makeatother


\makeatletter
	\newcommand{\numbereqbysec}{
		\@addtoreset{equation}{section}
		\def\theequation{\thesection.\arabic{equation}}
		}
\makeatother

\newenvironment{mptbl}{\begin{center}}{\end{center}}
\newenvironment{minipagetbl}[1]
	{\begin{center}\begin{minipage}{#1}
		\renewcommand{\footnoterule}{} \begin{mptbl}}%
	{\vspace{-.1in} \end{mptbl} \end{minipage} \end{center}}



\def\arcdeg{\hbox{$^\circ$}}


\newcommand\nonsequitor{\emph{non\ sequitor}}
\newcommand\apriori{\emph{a\ priori}}
\newcommand\Apriori{\emph{A\ priori}}
\newcommand\perse{\emph{per\ se}}
\newcommand\viceversa{\emph{vice\ versa}}
\newcommand\etal{\emph{et\ al}}
\newcommand\terrafirma{\emph{terra\ firma}}
\newcommand\adhoc{\emph{ad\ hoc}}
\newcommand\enroute{\emph{en\ route}}
\newcommand\ala{\emph{a\ la}}
\newcommand\opcit{\emph{op.\ cit.}}
\newcommand\ie{i.e.}
\newcommand\eg{e.g.}
\newcommand\vs{\emph{vs.}}


\newif\iffigavailable
	\def\figavailable{\figavailabletrue}
	\def\nofigavailable{\figavailablefalse}
	\figavailable
\newcommand{\Fig}[3]{
	\begin {figure} [t]
		\centering\leavevmode
		\iffigavailable\epsfbox {#1.eps}\fi
		\caption {#2}
		\label {f:#3}
	\end {figure}
}


\def\ednote#1{\noindent== \emph{#1} ==}		
\def\note#1{\noindent====\\\emph{#1}\\====}	
\def\revising{\ednote{to be revised}}
\def\comment#1{}			
\newcommand{\bold}[1]{\mathbf{#1}}
\newcommand\degree{$^\circ$}
\newcommand{\Qed}{$\bold{\Box}$}
\newcommand{\order}[1]{\times 10^{#1}}


\newcommand{\EQ}{\begin{equation}}
\newcommand{\EN}{\end{equation}}
\newcommand{\EQA}{\begin{eqnarray}}
\newcommand{\ENA}{\end{eqnarray}}
\newcommand{\LD}{\begin{description}}
\newcommand{\DE}{\end{description}}
\newcommand{\LI}{\begin{itemize}}
\newcommand{\LE}{\end{itemize}}
\newcommand{\LN}{\begin{enumerate}}
\newcommand{\NE}{\end{enumerate}}
\newcommand{\VB}{\begin{verbatim}}
\newcommand{\VE}{\end{verbatim}\\}
\newcommand{\QB}{\begin {quotation}}
\newcommand{\QE}{\end {quotation}}
\newcommand{\And}{\wedge}
\newcommand{\Or}{\vee}

\title {
	The correct analysis and explanation
	of the Pioneer-Galileo anomalies
}
\author {
\\
V Guruprasad\\
IBM T J Watson Research Center,\\
Yorktown Heights, NY 10598, USA.\\
}
\address{
prasad@watson.ibm.com
}
\date{1999.09.17}

\maketitle

\begin {abstract}

Tidal distension of spacecraft electronics 
due to spin and solar and galactic gravitation
elegantly explains \emph{all variations} in the anomaly
reported by Anderson \etal{}.
Contrary to their conclusion,
a constant residue seems to be present in
lunar, terrestrial, and possibly planetary measurements,
posing a problem of wider, more fundamental significance.

\end {abstract}
\vspace {1in}
\begin {center}
\begin {minipage} [b] {5in}
\begin {deflist} {Keywords}
\item[PACS]
95.55.Pe,	
95.30.Sf,	
96.35.Fs,	
95.10.Km	
\item[Keywords]
	Pioneer-10/11 anomaly, planetary ranging
\end {deflist}
\end {minipage}
\end {center}
\newcommand\s{s$^{-1}$}
\newcommand\ssq{s$^{-2}$}

\section {Introduction and overview}
\label {s:problem}

The Pioneer and other deep space missions
employing spin-stabilised spacecraft
have revealed an anomaly with characteristics
that have defied a simple explanation
\cite {Anderson1998}.
What has been measured is
a residual Doppler shift $\delta z$ in the telemetry signal,
which has been generally interpreted
as an acceleration by the gravitational redshift analogy
\EQ \label {eq:gaccel}
	\delta g \sim r^{-1} \, c^2 \, \delta z ,
\EN
$r$ being the distance from the sun.
Numerous mechanisms have been examined that
could yield this force,
including helium leakage and
the reaction from power dissipation and radiation
\cite {Murphy1998}
\cite {Katz1998},
but to no avail
\cite {Turyshev1999}
\cite {Anderson1999a}.
It has been noted that
the anomaly is of the same order as the Hubble constant
\cite {Rosales1998},
or the cosmological time dilation (CTD)
\cite {Ellman1998},
but the standard model prohibits
either on the planetary scale of distances:
the hypothesis of expansion on smaller scales
as such presents a logical difficulty
\cite [p.619] {MTW} and
matter appears to be gravitationally bound
all the way up to the galactic scale.
A modified theory of gravitation has also been considered
hypothesising a changeover from $r^{-2}$ to $r^{-1}$ character
that could produce this anomalous force,
but it fails to explain the \emph{absence} of the anomaly
in the planetary ranging data
\cite {Anderson1998}.
The Pioneers also display
an almost constant residual anomaly beyond $40$~AU,
but with a persistent difference
that again defies the equivalence principle
\cite {Anderson1998},
questioning the adequacy of any relativistic model whatsoever.
Furthermore,
even the residual value exhibits oscillations
that display synchronicity with the \emph{earth}'s orbital period,
though measurement and analysis errors
are stated to be ironed out. 

These experiences indicate that
the previous attempts have been simplistic
in seeking a single cause for the anomaly and
in viewing it as an actual acceleration,
given the lack of visual or quadrature evidence
to verify the inference.
The frequency shift is not sufficient to imply
a Doppler or gravitational cause,
as any means to cause the onboard tuning devices and circuits
to drop, or appear to drop, in frequency
would lead to the same observation.
Thermal and structural stresses can be presumably ruled out,
since they would have been already considered in
the design and the testing phases of the missions,
but not mechanisms
that would be ordinarily absent or too small
to be routinely dealt with on earth.
Such a mechanism cannot explain
the constant remaining anomaly of either Pioneer,
so a relativistic cause is still needed,
but the latter cannot be sufficient by itself
because of the violation of the equivalence principle, as mentioned.

Accordingly,
we must consider a combination of such causes,
thereby partitioning the anomaly into a constant part
of the order of the Hubble constant,
for which a relativistic explanation may yet be possible, and
a variable part,
whose path-dependence and oscillations,
together with the fact that
the anomaly has been observed only in spinning spacecraft,
suggest a connection to the gravitational tidal forces due to the spin.
This variable component is therefore attributed to
an actual, physical expansion of the onboard electronics,
in proportion to the tidal forces,
making the transponder response frequency decrease
without causing dynamical acceleration.
Its analysis exclusively concerns
the equivalent time dilation of the onboard clocks
\cite {Anderson1998},
which not only avoids the detracting notion of acceleration,
but is factually closer to the actually measured quantity $\delta z$.
It is convenient for this purpose, therefore,
to define a ``Hubble measure'' for the acceleration,
\EQ \label {eq:hmeasure}
	h = c^{-1} \, \delta g ,
\EN
having the same dimensions [$T^{-1}$] as the Hubble constant $H$,
which describes the anomaly as if it were CTD,
as explained in \S\ref{s:planetary}.
We thus have the decomposition
\EQ \label {eq:hbrkup}
	h = h_s (r, \bold{\hat{\omega}}) + h_c .
\EN
$h_s$ denoting the variable part due to tidal action,
$\bold{\omega}$, the spin, and
$h_c$, the constant part.

The possibility of tidal action stems from
the smallness of the anomaly,
$\sim 2.8 \order{-18}$~\s{}
\cite {Anderson1998},
taking about $30$~y to stretch the spacecraft by $1$~nm,
which explains why
the electronic circuits have continued to function,
albeit at gradually diminishing frequencies.
It also explains, as we shall see,
the proportionality of $\delta z$ to the expected tidal forces,
which requires a notion of linearity;
a larger rate of expansion would have caused
non-linear variations in the behaviour of the circuits,
disrupting their operation.
We ordinarily expect the molecular bonding forces
to prevent such a process,
but we do have the example of glasses
that ``flow'' slowly even under constant gravitational stress, and
the repetitive shear stress due to the spin should cause
a more rapid buildup of the microscopic fractures.
Rheological degradation is allowed for
in most component tolerances,
but is too small to be modelled in engineering studies and
very unlikely to have been considered in the spacecraft simulations.
The tidal buildup also explains
the manifestation of the anomaly in deep space,
where gravity is much weaker than on the earth's surface, and
seems to correctly fit the oscillatory pattern in
the Pioneer 10 anomaly.
In any case,
the possibility can be readily verified by spinning
a test spacecraft on earth in powered-on state and
continuously ``ranging'' it using telemetry.

Assuming these ideas to be correct,
we would expect $h_s$ would be maximum when 
the spin $\bold{\omega}$ is perpendicular to
the total gravitational force $\bold{g}$, and
should vanish while inline,
yielding three degrees of freedom in the model:
\EQ \label {eq:spin}
	h_s =	k_s (\phi) \, \omega^\alpha g^\beta \;
			| \bold{\hat{\omega} \wedge \hat{g}} | ,
\EN
where $\phi$ denotes the angle of rotation and
$k_s$, $\alpha$ and $\beta$ may vary between spacecraft
(Fig.\ \ref{f:Anom}).
The $\phi$-variation is expected
because of the mechanical anistropy due to
the rectangular layout of components
on electronic circuit boards (Fig.\ \ref{f:Struct}), and
is a prediction of the model.
The earth-synchronism of Pioneer 10 oscillations
is explained by the fact that
the spin axis $\bold{\hat{\omega}}$ is
generally made to follow the earth
for the telemetry to be possible at all
(Fig.\ \ref{f:Orbit}).
The persistent remaining difference in the Pioneer anomalies
seems to fit the asymmetric vector contribution
expected from the galactic gravitational field
(Fig.\ \ref{f:Galactic}).
A determination of $k_s (\phi)$, $\alpha$ and $\beta$
in the laboratory
would therefore provide a useful tool
for analysing the galactic gravitational field.
The detailed application of $h_s$
to the reported characteristics of the anomaly
is given in \S\ref{s:tidal}.

The absence of the anomaly in planetary data
is partly explained by the lack of a transponder 
to reproduce $h_s$ and its variations.
The constant portion $h_c$ presents some difficulty,
as it implies a uniform slowing of clocks, like the CTD,
in the planetary space,
questioning the standard model assumptions,
as already mentioned.
Its similarity to the CTD implies, however,
that $h_c$ could manifest as a Hubble flow,
\ie{} in proportion to the range,
rendering Anderson \etal{}'s consideration of
the inner planet ranging data insufficient for claiming
the absence of the anomaly in planetary motion,
especially given that
all six missions specifically involve $r > 1.3$~AU
\cite {Anderson1998}.
A search for Hubble flow-like evidence in
planetary, lunar and terrestrial data
does turn out to be positive
(\S\ref{s:planetary}).

A complete explanation of the anomaly is thus claimed
in the context of the deep space missions.
It is further concluded that
$h_c$ can and should be explained in a wider context, on grounds
that we are otherwise able to account for
\emph{all} of the discernible details in the NASA-JPL reports,
as explained in the next section, and
that the deep space ranging appears to be the first means
capable of verifying the absence of planetary CTD or Hubble flow
in the first place, and
may be indicative of new physics.

\section {Application of the model}
\label {s:tidal}

Accordingly,
the following principal characteristics of the anomaly
are sought to be explained by the model:
\LN
\def\theenumi{\Alph{enumi}}
\def\theenumii{\arabic{enumii}}

\item	Differs from planetary ranging
	\cite {Anderson1998}
	and between the spacecraft and missions
	\cite {Turyshev1999}:
	\LN
	\item	Does not manifest in planetary ranging.
		\label{i:noplanet}

	\item	Earth clock drift fits Doppler but not range.
		\label{i:badrange}

	\item	Fitting a line-of-sight acceleration leads to
		sign inconsistency.
		\label{i:badspace}

	\item	Speed of light correction to the planetary forces
		does not fit Galileo.
		\label{i:gspeed}
	\NE

	(\ref{i:noplanet}) has been partly explained by the fact
	that the ranging procedure is indeed
	different from that of planetary ranging,
	as it employs a generated return and
	not an actual reflection.
	Completeness of (\ref{i:noplanet}) is questioned
	since only inner planet range data has been used
	in arriving at the conclusion,
	whereas the spacecraft displaying the anomaly
	are invariably at $r > 1.3$~AU and generally
	about or beyond the outer planet distances.

	(\ref{i:badrange}-\ref{i:gspeed})
	are the reasons for the present view
	that the variable part $h_s$
	cannot mean actual acceleration and
	must be explained as
	a slowing down of the \emph{spacecraft} clocks,
	which can depend on
	the composition and motion parameters of
	the individual spacecraft.
\label {i:overall}

\item	Fluctuations and periodicities
		\cite {Turyshev1999}:
	\LN
	\item	Sinusoidal variation with earth-year periodicity
		(Pioneer 10: 1987-1994).
		\label{i:annual}

	\item	Maxima larger than minima,
		data points wilder near the minima
		(Pioneer 10).
		\label{i:asymmetric}

	\item	More fluctuations later
		(Pioneer 10: mid-year after 1993,
		more frequent after 1996).
		\label{i:fluct}
	\NE

	These cannot be explained by modified
	gravitational field theories
	\cite {Anderson1998}, and
	force-generating mechanisms like leakage and radiation
	have already been ruled out as unlikely,
	along with systematic observation error
	that might be inferred from (\ref{i:annual})
	\cite {Turyshev1999}.
	These features are well explained in the present model by
	the targeting of the spin axis $\bold{\hat{\omega}}$
	towards the earth.
\label {i:osc}

\item	Range dependence:
	\LN
	\item	Higher in nearer orbits (Ulysses \vs{} Pioneers),
		not matching $r^{-2}$
		\cite {Anderson1998}.
		\label{i:perihelion}

	\item	Linear weakening from Jupiter
		to somewhere $> 40$~AU
		(Pioneer 10: 1987-1994)
		\cite {Turyshev1999}.
		\label{i:linear}

	\item	Eventually levels off (both Pioneers)
		\cite {Anderson1998}
		\cite {Turyshev1999}.
		\label{i:residual}

	\item	Continues to differ between the two Pioneers
		as they head out of the solar system
		\cite {Anderson1998}
		\cite {Turyshev1999}.
		\label{i:galactic}

	\item	Remains larger than $H$, at $\sim 84$~km/s-Mpc.
		\label{i:expanding}
	\NE

	(\ref{i:expanding}) parallels other data
	that seem to confirm the general occurrence of
	$h_c \approx H$ on the planetary scale,
	to be discussed in \S\ref{s:planetary}.
\label {i:range}
\NE

Fig.\ \ref{f:Anom} illustrates
the basic quantities in the model (eq.\ \ref{eq:spin}).
Each spacecraft is approximately cylindrical, and
carries the main antenna on one end,
which is generally kept pointing towards the earth
by spin stabilisation and correcting manoeuvres.
For this,
the spin axis must point towards the earth,
so that it subtends the angle $\theta$
to the net gravitational force $\bold{g}$,
presumably due to the sun.
As the spacecraft moves, the value of $\theta$ changes,
causing $h_s$ to vary even if $\omega$ were constant,
as likely in the closed orbit missions, and
differently from the $r^{-2}$ nature of $\bold{g}$.
Additionally,
the construction must invariably result in a mechanical anisotropy,
illustrated in Fig.\ \ref{f:Struct},
which should cause $h_s$ to vary with the spin rotation angle $\phi$,
tracing out a hysterisis loop along each of
the tranverse principal axes.
The Pioneer 10 plots given by Turyshev \etal{}
show only $50$~day averages
\cite {Turyshev1999},
but continuous telemetry may have occurred
during Galileo's earth fly-by, and
might reveal high frequency oscillations due to this anisotropy.

Fig.\ \ref{f:Orbit} illustrates
the orbital variation of $\theta$
that explains (\ref{i:annual})-(\ref{i:fluct}),
(\ref{i:perihelion}), (\ref{i:linear}) and (\ref{i:galactic}).
$S$ denotes the sun and $A$, $B$, $C$ and $D$,
successive positions of the earth along its orbit.
If the spacecraft stays in the plane of the ecliptic,
as indicated by point $P_0$,
its spin axis would be pointing directly towards the sun
when the earth is at $A$ or $C$,
so that $\theta = 0$ and
there would be no $h_s$ contribution at these times.
$|\theta|$ reaches its maximum
$\theta_0 = \angle B P_0 S = \angle D P_0 S$,
when the earth is at $B$ or $D$,
so that $h_s$ should peak twice a year.
In the case of non-ecliptic spacecraft,
as indicated by point $P_1$,
$\theta$ would again have two maxima, $\theta_1$ and $\theta_2$,
but they would be now unequal and
occur at different times, $A$ and $C$.
At points inbetween $P_0$ and $P_1$, however,
the spacecraft would be seasonally occluded by the sun
when the earth is near $A$, in which case,
earth-targeting manoeuvres and measurements are both impossible.
This difference between ecliptic and non-ecliptic paths
may be responsible for the sign difference
in (\ref{i:badspace}).

The model clearly has adequate degrees of freedom
to explain the details of the anomaly.
For instance,
it is about $400 \order{-18}$~\s{}
($12 \order{-8}$~cm/\ssq{})
for Ulysses,
which orbits the sun at $1.3 < r < 5.2$~AU, and
is larger than in the case of the Pioneers.
(\ref{i:perihelion}).
As $\theta$ is small at $r \gg 1$~AU,
it can contribute a linear ($r^{-1}$) variation.
For example,
between $5$~AU (Jupiter) and $40$~AU (Neptune and Pluto),
$\max(\theta)$ drops from $11.32\arcdeg$ to $1.43\arcdeg$,
yielding a factor of $7.8 \approx 40/5$ in $\sin(\theta)$ and
thence in $h_s$, all other factors remaining the same.
In the case of Pioneer 10,
the spin rate $\omega$ was also decreasing upto mid-1990
(\ref{i:linear}), and
$\alpha$ is very likely to be $1$,
the $\theta$ variation cannot be the sole cause of this decrease.
In any case,
$\beta$ is likely to be fractional, and
the smallness of $|k_s|$ relative to $h_c$ may be sufficient
to mask out non-linear details.
Additionally,
since $|\sin(\theta)|$ is at most $1$,
the anomaly is rendered incapable of rising much further
at perihelions closer than $1$~AU to the sun.

(\ref{i:residual}) and (\ref{i:galactic})
are elegantly explained by considering
the vector sum of the solar and the galactic gravitational forces
$\bold{g} = \bold{g}_s + \bold{g}_g$
acting on the spacecraft
(Fig.\ \ref{f:Galactic}).
As the spacecraft are headed in opposite directions from the sun,
their spins must be approximately in line with the sun,
at least so long as their signals were being received.
Had their headings been also perpendicular to the galactic centre,
$\bold{g}$ would have been symmetrical
with respect to the sun, and
the expansion rates would have been equal in that case.
Pioneer 10 is, however, headed in the direction of the galactic centre,
so the vector sum $\bold{g}$ is
neither equal nor symmetric about the spin axis.
The galactic component $\bold{g_g}$ is shown greatly exaggerated:
in the limit of small $\bold{g_g}$,
where $\theta \approx \pi/4$,
the magnitude of the sum, $|\bold{g}|$,
varies more rapidly than $\sin (\theta)$,
explaining the larger anomaly of Pioneer 11.
This not only explains the disparity,
but confirms that the spacecraft are indeed still expanding,
since $h_c$ may not be due to expansion.

As remarked by Turyshev \etal{},
the spin deceleration of $4.73 \order{-6}$~\ssq{}
($0.065$~rev/day/day)
would have contributed to the linear decrease
seen in the Pioneer 10 graphs
\cite {Turyshev1999},
but we do not have the data to determine $\omega$ at any point,
in order to determine the exact contribution,
from which $\beta$ could have been determined,
assuming $k_s(\phi) \approx \alpha = 1$.
The absence of a semi-annual maximum upto mid-1996
(\ref{i:annual})
is generally explained by occlusion by the sun
(point $A$ in Fig.\ \ref{f:Orbit}),
given that Pioneer 10 has not been too far from the ecliptic.
Being cumulative,
(eq.\ 1 in \cite {Turyshev1999}),
the anomaly would continue to grow at the rate set 
by the last targeting manoeuvre during the occlusion.
We do not know
how frequently the targeting manoeuvres were being performed,
but it seems less likely
that they would have been needed from near $A$,
where $\theta$ changes the least,
than from points $E$ and $F$,
before and after the occlusion, respectively, and
is most likely around $C$,
where $\theta$ varies most rapidly.
This would explain why the maxima are larger, and
the lower data points are more scattered 
(\ref{i:asymmetric})
because the spin is allowed to drift on the side of $A$.
There are a few data points very close to the best-fit minima
(mid-1987, mid-1988),
which at first sight seem to contradict the occlusion idea,
but these too can be attributed to
the cumulative nature of the anomaly.

Changes in the spin rate have clearly contributed
to the variations between mid-1990 and mid-1992.
The gradual appearance of more humps thereafter
is probably due to drift of the spin axis,
as a re-targeting was needed in Jan 1997.
The spin has been otherwise steady,
explaining the levelling out of the anomaly
(\ref{i:residual}).

\section {Constant part of the anomaly}
\label {s:planetary}

As already remarked,
the ranging data from the two nearest planets Venus and Mars
\cite {Anderson1998}
is not sufficient for concluding that
the (constant part $h_c$ of the) anomaly
is absent in the planetary data, because
\LN
\def\theenumi{\alph{enumi}}

\item	the measured anomalous quantity is
	the frequency shift $\delta z$,
	not an actual acceleration unambiguously
	determined by visual or quadrature means;
	\label {i:isdz}

\item	interpretation as actual acceleration does not fit the data
	(\ref{i:badrange}-\ref{i:gspeed}),
	regardless of whether it is
	ascribed to leaking and dissipation or
	to modified gravity;
	\label {i:accnofit}
and
\item	the anomaly has been observed exclusively in deep space
	beyond Mars, not in near space missions.
	\label {i:deepspace}

\NE
Items (\ref{i:isdz}) and (\ref{i:accnofit}) particularly imply
that the true nature of the anomaly lies only in
the spacecraft time dilation, or
an equivalent expansion of onboard clocks,
which is how the tidal expansion $h_s$ was deduced.
The expansion of clocks does not imply
a uniform expansion of space, as $h_s$ itself illustrates,
but a uniform expansion, if present,
would include the expansion of clocks, yielding a redshift.
The associated Hubble's law recession
\EQ \label {eq:hubble}
	\dot{r}_c = h_c \, r
\EN
implies an increasing optical path $r \sim c t$, hence,
if we did not know the cause,
we would only observe the apparent acceleration
\EQ \label {eq:accel}
	\ddot{r} = \frac{d}{dt} \{ h_c \, c t \} = h_c \, c ,
\EN
equivalent to eq.\ (\ref{eq:hmeasure}),
in this case.
This is substantially the derivation used by Rosales and Sanchez-Gomez
\cite {Rosales1998}.

Anderson \etal{} looked for a radial acceleration,
which would have measurably perturbed
the orbital radius and period
\cite {Anderson1998},
but this is insufficient in the context,
as the anomaly could have been caused,
according to eq.\ (\ref{eq:accel}),
by an unmodelled recession instead of static perturbations.
The absence of recession cannot be verified by a one-time observation
because of the smallness of $h_c \approx H$
(\ref{i:expanding}):
the expectable recession velocities would be only
$1.6$~$\mu$m/s for Jupiter $5$~AU) and
$12.5$~$\mu$m/s for Neptune/Pluto ($40$~AU),
too small to be detected by Doppler measurement.
The only way to detect the ongoing recession, if any,
is to look for cumulative displacements
over a long enough period, say a year,
as listed in Table \ref{t:planetary} below.
The fact that
the standard model does not support this possibility
is not relevant to this test, in part
because this notion of the standard model
has itself never been empirically verified,
$H$ being too small to have inadvertently shown up
on any scale short of the galactic.

The table further shows
why it is incorrect to compare with Venus or Mars -
even their annual recessions with respect to earth
would be only about $2$-$6$~m,
less than a twentieth of the precision, $100$-$150$~m,
available for the earth and Mars orbital radii
from the Viking data
\cite {Anderson1998}.
One would ordinarily expect that by now,
the measurements would have been repeated,
with a gap of at least three years,
so that the recession, if present,
would have been noticed already.
Our research reveals that
such differences have been reported at least twice,
once for Mercury,
but were treated both times as systematic error.
The recent instance is finding
Jupiter $11$~km farther, in 1992, than predicted,
presumably from data acquired in the 1970s
\cite {Harmon1994},
indicating a recession of $611$~m/y, or $H \approx 770$~km/s-Mpc,
about $10$ times the possible Hubble flow.
Incidentally,
due to its angular limit of resolution,
the ranging precision follows a ``Hubble's law'' of its own
\EQ \label {eq:hmeas}
	\delta r \sim h_{\mu} \, r ,
\EN
where $h_{\mu}$ is characteristic of the measuring technique.
As a result,
the presence or absence of Hubble flow
can be decided only by improving the sensitivity
by at least two orders of magnitude.
Since this is precisely the capability in effect
achieved by the spacecraft ranging technique
used in the deep space missions,
the indication of $h_c$ may verily mean
the presence of a planetary Hubble flow.

It is of paramount importance, therefore,
to look for it at even smaller scales,
using more precise means
that are generally available because of proximity.
The logical difficulty referred to earlier should prevent
the flow from manifesting on the scale of
the internal structure of matter,
but since physics cannot rest on logic alone,
this too needs to be empirically verified.
We do find positive indications 
on lunar and geophysical scales, as follows.

The moon is known, from very precise measurements,
to be receding at $3.84$~cm/y
\cite {Lunar1994},
which is large enough to accomodate the anomaly along
with the known tidal cause for the recession,
there being no independent means of measuring
the tidal contribution.
The tidal action is also responsible for
the slowing down of earth's rotation, and remarkably,
this and other geological evidence have indicated that
the earth may have been expanding in the past 
\cite {Wesson1973}
\cite {Creer1965}
\cite {Runcorn1965}
\cite {Wesson1999}.
The indictated values of $0.4$-$0.6$~mm/y
correspond to the range $61$-$92$~km/s-Mpc for $H$
\cite {MacDougall1963},
of which a small part could still account for the tidal action,
so that the anomaly cannot be easily ruled out even in this case.

\section {Conclusion}
\label {s:concl}

We have essentially established that
there could be two distinct causes underlying the anomaly,
one of which ($h_s$) is associable with
an actual stretching of the spacecraft and its electronics
due to gravitational tidal action from its spin, and
is dependent on the individual spacecraft and its motion.
The presence of $h_s$ may be responsible for
the contradictory results obtained
when various models to explain the constant acceleration ($h_c$)
are applied to the data.
It has been further shown
that $h_s$ fits the anomaly in greater detail
than any of the uniform gravitational and
other force models previously attempted.

It has been also shown that
Anderson \etal{}'s conclusion
that $h_c$ is absent in planetary data, is premature,
as a uniform expansion of space could cause
the time dilation and give the appearance of acceleration,
without showing up even cumulatively in planetary ranging data,
let alone manifest in the Doppler residuals.
The possibility appears to be no more offbeat or remote than
the modifications to gravitation theory they have considered.
More importantly, our analysis shows
that the anomaly may itself be the first tool
precise enough to measure
the planetary flow and the galactic gravitation, and
may yet be pointing to significant new physics.

An illustrative potential application is
the resolution of Arp's quasars and galaxies
that exhibit redshifts in excess of the Hubble flow and
in quantised increments
\cite {Arp1988}.
Every such object appears to be
(a) spinning and (b) in intimate proximity of
a much larger galactic body,
like the Galileo and Pioneer spacecraft in relation to the sun,
suggesting that $h_s$ may be responsible for these anomalies as well.

\acknowledgements

Special thanks are owed to
Bruce Elmegreen, A Joseph Hoane and others here at IBM Research,
for discussions during this work, and
to Alaign Milsztajn for a numerical correction
that strengthened the case.


\clearpage

\Fig{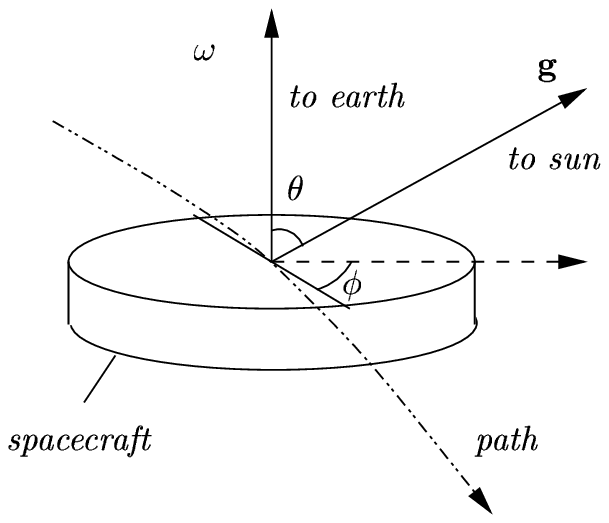}	{Geometry of the anomaly}		{Anom}
\Fig{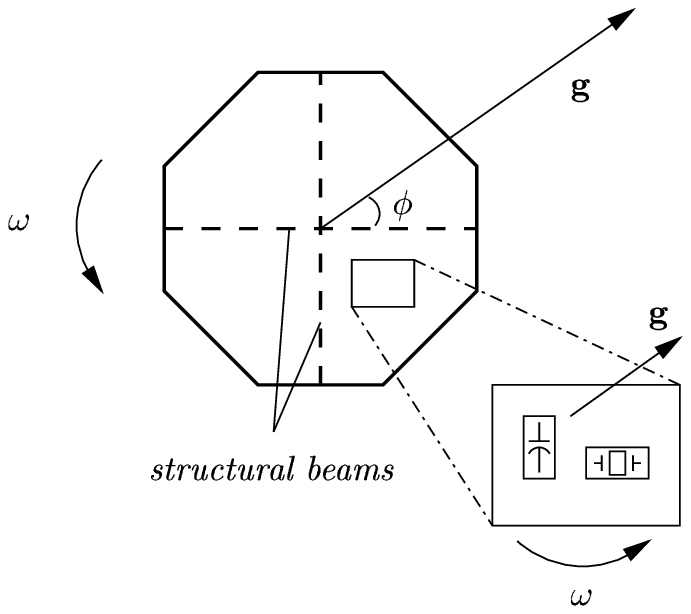}	{Spacecraft construction schematic}	{Struct}
\Fig{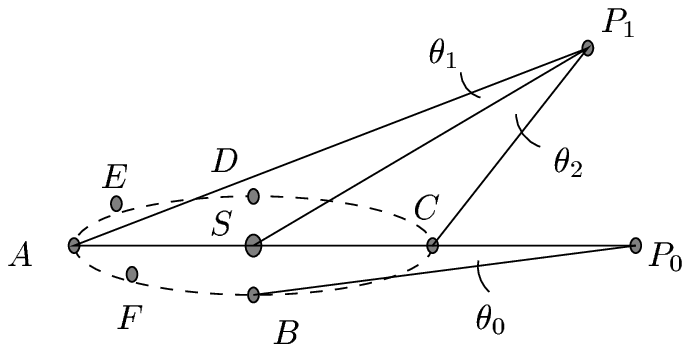}	{Orbital geometry}			{Orbit}
\Fig{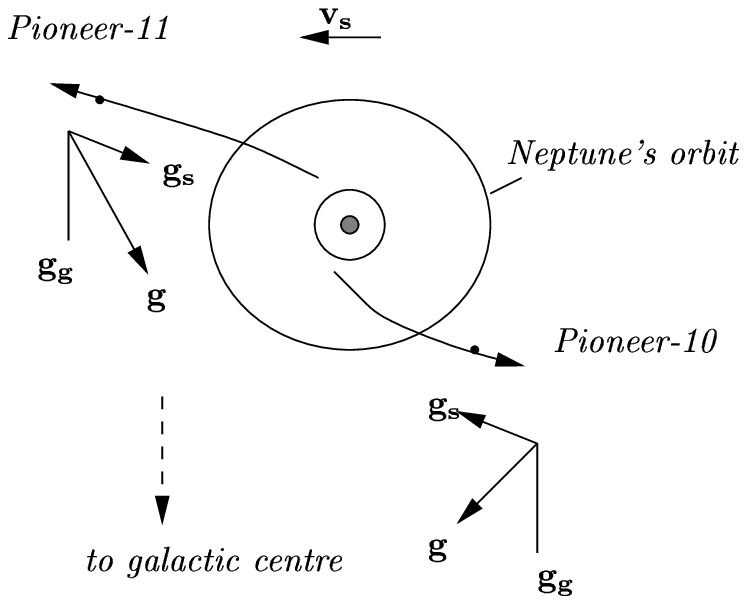}	{Galactic contribution geometry}	{Galactic}

\begin {table}[htbp]
\begin {minipagetbl}{3.25in}

\begin {tabular}{|| l | r@{}r@{}l@{} | r@{}r@{}l@{} | r@{}r@{}l@{} ||}
\hline
\emph{Body}
& \multicolumn{3}{c|} {\emph{Distance}}
& \multicolumn{6}{c||}{\emph{Recession} (m/y)}	\\
& \multicolumn{3}{c|} {(AU)}
& \multicolumn{3}{c}  {$h_c = 50^\dag$}
& \multicolumn{3}{c||}{$h_c = 75^\dag$}				\\
\hline
Moon$^\ddag$
	& $2.57 \times$&&$10^{-4}$
		& 	0.0197&&	&	0.0295&&\  \\
\hline
Mercury	&	0&.&40	&	30&&		&	44&&	\\
Venus	&	0&.&72	&	55&&		&	83&& 	\\
Earth	&	1&.&00	&	77&&		&	115&& 	\\
Mars	&	1&.&52	&	117&&		&	175&& 	\\
\hline
Jupiter	&	5&.&20	&	398&&		&	597&& 	\\
Saturn	&	9&.&56	&	731&&		&	1,096&&	\\
Uranus	&	19&.&19	&	1,468&&		&	2,202&&	\\
Neptune	&	30&.&11	&	2,303&&		&	3,455&&	\\
Pluto	&	39&.&53	&	3,024&&		&	4,535&&	\\
\hline
\end {tabular}
\footnotetext[0]{$^\dag$ in km/s-Mpc,
	$= 1.62, 2.43\order{-18}$~\s{}, respectively. }
\footnotetext[0]{$^\ddag$ from earth; measured: $0.0384$~m/y (1994)}
\end {minipagetbl}

\caption {Planetary Hubble flow}
\label {t:planetary}
\end {table}

\end {document}